\documentclass{kapproc} % Computer Modern font calls
\usepackage{graphicx}
\newcommand{\Gray}{$\gamma$-ray\ }
\newcommand{\Grays}{$\gamma$-rays\ }
\begin{document}

\articletitle[TeV Observations of SNRs and unidentified sources]
{TeV Observations of SNRs and \\ unidentified sources}

\author{Stephen Fegan}
\affil{Fred Lawerence Whipple Observatory\\
Harvard-Smithsonian CfA\\
P.O. Box 97, Amado, AZ 85645-0097, U.S.A.\\
\vspace*{1ex}\\
Department of Physics\\
University of Arizona\\
1118 E. 4th Street, Tucson, AZ 85721, U.S.A.}
\email{sfegan@egret.sao.arizona.edu}


\begin{abstract}

A review of very high energy \Gray astronomy is presented. Particular
attention is paid to the atmospheric Cherenkov imaging technique whose
employment has resulted in detections of both galactic and
extra-galactic objects at energies above 300GeV. Next generation
ground-based telescopes promise to increase our knowledge of extreme
astrophysical objects as they begin to operate over the next few
years.
\end{abstract}

\section{Introduction}

Very High Energy (VHE) \Gray astronomy, or TeV \Gray astronomy as it
is sometimes known, traditionally describes observations in the energy
range from 300GeV to 100TeV. Doing astronomy in this energy range was
made possible by the development of the atmospheric Cherenkov imaging
technique, which led to the detection of the first VHE source, the
Crab Nebula, more than ten years ago (Weekes et al. 1989). Ground-based 
instruments operating in this energy region typically have large
collecting areas, high angular resolution and good energy resolution
(Ong 1998).

In the years since the detection of the Crab Nebula many other
galactic and extra-galactic VHE \Gray sources have been detected.
There are observatories operating in the northern and southern
hemispheres which have been able to independently confirm a number of
the claimed detections. Ground-based telescopes have participated in
multi-wavelength observations with satellite-based detectors, yielding
many interesting results. To date only 1\% of the VHE sky has been
observed at these energies.

The coming years will bring a new generation of ground-based VHE \Gray
observatories that will build on the techniques developed over the
past three decades. A number of large, next generation, Cherenkov
telescopes are either starting to operate or being built around the
world. The first decade of the new millennium promises many new
exciting results in this field.

\section{Ground-Based Telescopes}

At very high energies, \Grays and cosmic-rays interact with the
atmosphere to produce Extensive Air Showers (EAS), showers of charged
secondary particles which propagate through the atmosphere, reaching
the ground if the energy of the primary particle is sufficiently high.
Detecting these showers and inferring the composition, energy and
arrival direction of the primary particle is the challenge of
ground-based VHE astronomy.  Fortunately, the mechanism by which the
shower is produced is very well known and can be modeled for
different primaries at any energy.

Directly detecting the secondary particles is possible for instruments
placed at high altitude and for sufficiently energetic primaries. For
lower energy primaries, the cascade of charged particles does not
reach the ground but can be indirectly detected from the Cherenkov
radiation emitted as the relativistic charged particles traverse the
atmosphere at speeds in excess of the speed of light in air. The
Cherenkov photons are strongly beamed in the arrival direction of the
incident photon and form a narrow cone of light as they travel down
through the atmosphere. A telescope with fast light detectors can
sample the $\sim5\textrm{ns}$ pulse of Cherenkov light. The total
number of photons recorded is roughly proportional to the energy of
the primary, unless the telescope samples at the edge of the shower,
where the number of photons falls off quickly.

Historically, the most successful ground-based detection method has
been the Imaging Atmospheric Cherenkov Technique originally proposed
in 1977 (Weekes \& Turver 1977). A large telescope collects Cherenkov
light from showers and focuses it onto a plane of photomultiplier
tubes. Imaging the shower in such a fashion produces a ``picture'' of
its development through the atmosphere. Typically, these images are
characterized by a number of straight forward parameters (Hillas 1985)
and it is on the basis of subtle differences in these parameters that
hadronically produced showers can be differentiated from purely
electromagnetic showers and rejection of cosmic-ray particles can be
made. Although the isotropic background consists of $\sim1000$ times
more cosmic-rays than $\gamma$-rays, a single imaging atmospheric
Cherenkov telescope (IACT) can reject $~99.7\%$ of the events caused by
these background showers, leaving a statistically detectable
\Gray signal, if one is present and sufficiently strong. An array of
such telescopes working in coincidence can reject substantially more
(Daum et al. 1997).

Since the detection of the Crab Nebula in 1989, VHE gamma-ray
astronomy has advanced in two major ways. First, the development of
high resolution cameras has allowed better imaging of the showers and
more refined rejection of hadronic background events. A number of
observatories such as the Whipple 10m telescope, CAT, a French
telescope in the Pyren\'{e}es (Barrau et al. 1998) and CANGAROO, a
Japanese-Australian telescope in Woomera, Australia (Hara et al. 1993)
employ this approach. The second major development is the combination
of a number of IACTs to form an array that can view the shower
development from a number of different points on the ground. Such
stereoscopic viewing increases the suppression of hadronic background
and allows for better reconstruction of the arrival direction and
energy of the primary photon. To date the most significant example of
such an array is HEGRA, five small imaging telescopes on La Palma run
by an Armenian-German-Spanish collaboration (Daum et al. 1997).

\begin{figure}[ht]
\centerline{\resizebox{!}{1.8in}{\includegraphics[]{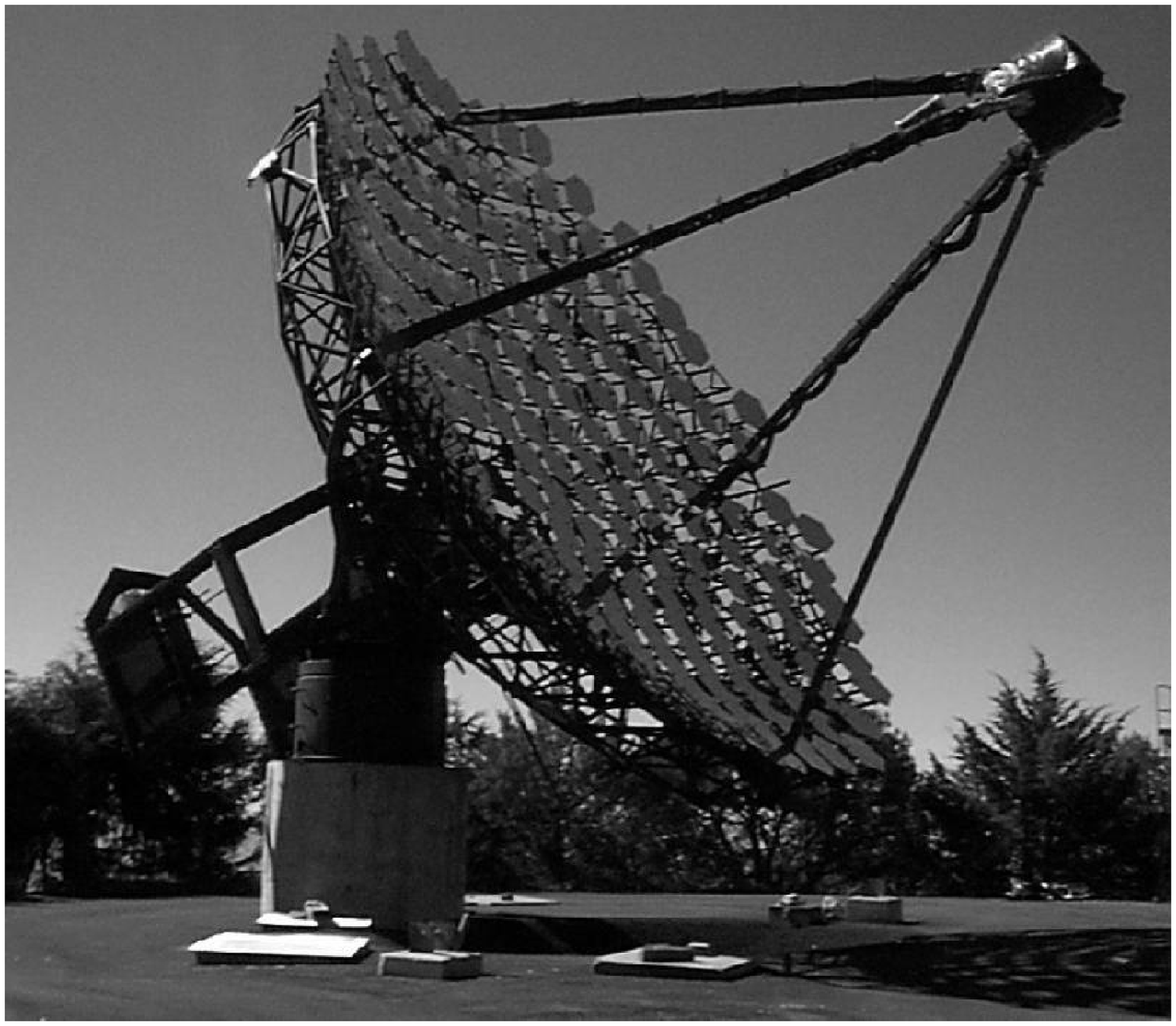}}
            \resizebox{!}{1.8in}{\includegraphics[]{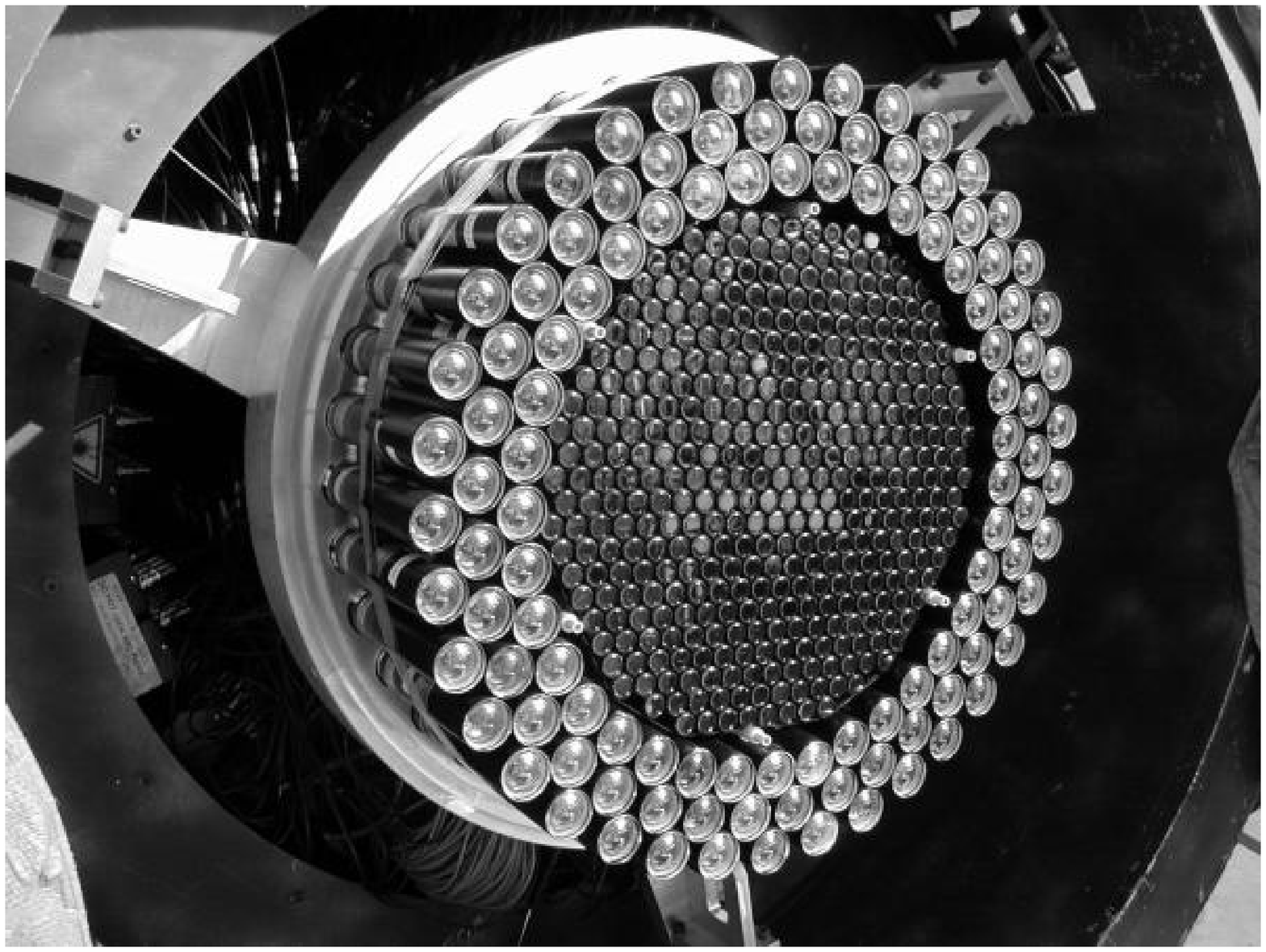}}}

\caption{\label{GBT::WHIPPLE}Left, the Whipple observatory 10m 
imaging atmospheric Cherenkov telescope. Right, at the focus, a 490
pixel high resolution camera operating since 1999.}
\end{figure}

\section{Supernova Remnants}

\subsection{Plerions}

The Crab Nebula has been detected in VHE \Grays by many independent
Cherenkov observatories in the northern hemisphere and by one in the
southern. Routine observations of the Crab Nebula by IACTs continue
and it has come to be regarded by the VHE community as the
``standard candle'' and is often used to provide a check on the
calibration of new instruments. Modern IACTs can detect the Crab
Nebula at the 5-6$\sigma$ level after one an hour of observations, a
testament to how far the field has advanced since the original
detection in 1989, which was a 9$\sigma$ excess based on 60 hours of
observations. Recently the STACEE and CELESTE collaborations, have
published significant detections of the Crab Nebula in the energy
region $E>190\pm60\mathrm{GeV}$ (Oser et al. 2000) and
$E>50\mathrm{GeV}$ (De Naurois et al,. 2000) respectively, an energy
range that extends lower than other ground-based instruments. At the
other end of the VHE spectrum, the Crab was the first \Gray source
source detected by an air shower array, when it was seen by the Tibet
Air Shower Array at $E>3\mathrm{TeV}$ (Amenomori 1999).

The energy spectrum between 300GeV and 50TeV has been well established
by a number of groups (Hillas et al. 1998, Aharonian et al. 2000). To
date no pulsation has been seen in the VHE \Gray signal from the Crab
(Gillanders et al. 1997; Burdett et al. 1999; Aharonian et al. 1999).
The VHE flux is thought to arise, in the most part, from inverse
Compton scattering of synchrotron photons by relativistic electrons,
so called synchrotron self-Compton or SSC emission (De Jager \&
Harding 1992; Hillas et al. 1998). By fitting the SSC model to the
energy spectrum that results from combining the X-ray, EGRET and VHE
data, a magnetic field of $160\mu G$ is derived (Hillas et al. 1998).

\begin{table}[ht]
\caption[TeV Observations of Plerions]
{\label{PLERIONS::OBSTABLE}TeV Observations of Plerions}
\begin{tabular*}{\textwidth}{@{\extracolsep{\fill}}lcc}\hline
{\bf Object} & {\bf Exposure time} & {\bf Flux/Upper Limit} \\
{\bf Name} &{\bf (hours)} &{\bf  x $10^{-11}cm^{-2}s^{-1}$} \\\hline

{\bf  EVERYONE} & & \\
Crab Nebula & $\rightarrow\infty$ & 7.0 ($>$ 400GeV) \\
{\bf  CANGAROO} & & \\
PSR 1706-44 & 60 & 0.15 ($>$1TeV) \\
Vela Pulsar & 116 & 0.26 (E/2 TeV)$^{-2.4}$ TeV$^{-1}$  \\
{\bf  Durham} & & \\
PSR 1706-44 & 10 & 1.2 ($>$300GeV) \\
Vela Pulsar & 8.75 & $<$5.0 ($>$300GeV) \\
\hline
\end{tabular*}
\end{table}

No other plerions have been detected by IACTs in the northern
hemisphere; however the situation is different in the southern
hemisphere where the CANGAROO group have detected two. In 1993 they
reported the detection of PSR 1706-44 based on 60 hours of
observations (Kifune et al. 1995). Confirmation by the Durham group
based on 10 hours of observation was subsequently made (Chadwick et
al. 1997). The VHE source is associated with a pulsar of period 102ms
and appears to be associated with a supernova remnant. No pulsations
have been detected in the VHE data. The CANGAROO group has also
reported the detection of a VHE signal in the neighborhood of the Vela
pulsar. The detection is at the $6\sigma$ level based on $\sim120$
hours of observation and the measured flux above 2.5TeV is $0.25\times
10^{-11}
\mathrm{cm}^{-2}\mathrm{s}^{-1}$.  The VHE signal, which is offset
from the location of the pulsar by $0.14^\circ$, is thought to
originate from a synchrotron nebula, powered by a population of
relativistic electrons which were created in the supernova explosion
and which have survived since then due to the low magnetic field in the
nebula. The nebula is assumed to be centered on the birthplace of the
pulsar, which was ejected at the time of the supernova explosion.

\begin{figure}[ht]
\centerline{\resizebox{!}{2.5in}{\includegraphics[]{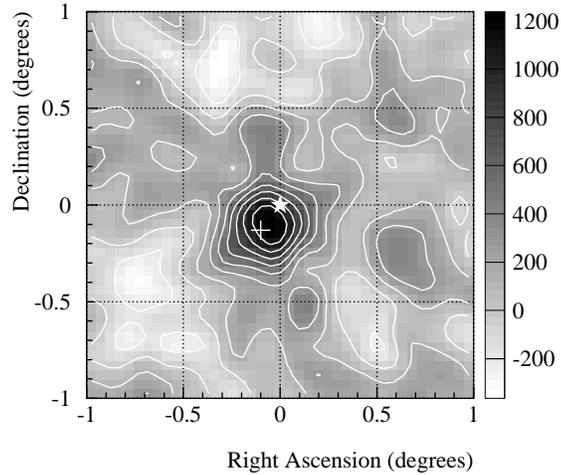}}}

\caption{\label{PLERIONS::VELA}VHE emission in the neighborhood of the 
Vela Pulsar as detected by the CANGAROO experiment. The contours show
the number of excess events per deg$^2$. The present location of the
pulsar is marked with a star.}
\end{figure}

\subsection{Shell-type}

It has long been hoped that VHE \Gray astronomy could provide a probe
of the origin of cosmic-rays with E$<10^{14}$eV. Supernovae are
regarded as the most likely producers of these cosmic-rays as they are
the only galactic objects capable of supplying the power required to
account for the observed cosmic-ray energy spectrum. Furthermore,
diffusive shock acceleration (Blanford \& Ostriker 1978, Bell 1978)
provides a natural mechanism to convert the kinetic energy of the SNR
shock-front into a spectrum of accelerated charged particles with
dN/dE=E$^{-2.1}$. This source spectrum, after correcting for diffusion
in the galaxy (Swordy et al. 1990), fits the locally observed spectrum
of E$^{-2.7}$. If this mechanism is correct, then interactions between
the relativistic charged particles and the interstellar medium
surrounding the SNRs should produce \Grays through the decay of
secondary $\pi^0$ particles (Drury, Aharonian \& V\"{o}lk 1994 --
DAV). These models predict that fluxes of \Grays should be high enough
to be visible to the current generation of satellite-based and
ground-based detectors. Detection of the signature $\pi^0$ bump at MeV
energies and a spectrum extending to tens of TeV would be a clear
indication that cosmic-ray acceleration does take place in
SNRs. However, the experimental situation is complicated by the
presence of a population of \Grays produced by the inverse-Compton
reaction of relativistic electrons and the cosmic microwave
background. Separating these two components requires that the spectrum
be measured continuously from 10 MeV to 10 TeV.

For most of its history, VHE \Gray astronomy provided no detections of
shell-type supernova remnants. Observations of SNRs that are
considered to be good candidates for neutral pion decay have been
undertaken by numerous groups. In particular, observations of W44,
W51, $\gamma$-Cygni, W63 and Tycho's SNR, selected due to their
possible association with molecular clouds which should provide an
enhanced target for $\pi^0$ decay, their possible associations with
EGRET sources and their small angular extent (Buckley et al. 1998),
have failed to produce detections in VHE \Grays (Table
\ref{STSNR::OBSTABLE}). Buckley et al. 1998 suggest that, assuming the
EGRET emission is from these shell-type SNRs and that $\pi^0$ decay
dominates, the spectra of cosmic-rays produced would have to be softer
than suggested by DAV. The required differential source spectrum is
E$^{-2.5}$ for $\gamma$-Cygni and E$^{-2.4}$ for IC443. Even if no
reference is made to the EGRET data, the VHE upper-limits for some of
the objects (IC443) push the allowable parameter space of DAV.
Gaisser et al. (1998) performed fits of the EGRET and Whipple results
concluding that the EGRET data must be dominated at lower energies by
electron bremsstrahlung but the source spectrum must still be steep to
account for the Whipple upper limits.

\begin{table}[ht]
\caption[TeV Observations of Shell-type SNR]
{\label{STSNR::OBSTABLE} TeV Observations of Shell-type SNR}
\begin{tabular*}{\textwidth}{@{\extracolsep{\fill}}lcc}\hline
{\bf Object} & {\bf Exposure time} & {\bf Flux/Upper Limit} \\
{\bf Name} & {\bf (hours)} & {\bf x $10^{-11}cm^{-2}s^{-1}$} \\\hline
{\bf CANGAROO} & & \\
%Vela Pulsar{\it {\it $^1$}} & 116 & 0.26 (E/2 TeV)$^{-2.4}$ TeV$^{-1}$  \\
RXJ 1713.7-3946 & 66 & 0.53 ($\ge$1.8 TeV) \\
SN1006  & 34 & 0.46 ($\ge$1.7 TeV) \\
W28 & 58 & $<$0.88 ($>$ 5 TeV{\it $^a$} )\\
%Cen X3{\it $^3$} & 17 & $<$0.52 ($>$2 TeV) (\it{prelim})\\
%Vela X1{\it $^3$} & 12 & $<$0.55 ($>$2 TeV) (\it{prelim})\\
{\bf HEGRA} & & \\
%Cas A & $\sim232$ & 0.058$\pm$0.012$\pm$0.020 ($>$ 1 TeV) {\it $^b$} \\
Cas A & 232 & 0.058 ($>$ 1 TeV) {\it $^b$} \\
%Tycho & 37.6 & none \\
$\gamma$-Cygni & 47 & $<$1.1 ($>$500GeV){\it $^c$} \\
{\bf Durham} & & \\
SN1006 & 41 & $<$1.7 ($>$300GeV) \\
{\bf Whipple} & & \\
Monoceros & 13.1 & $<$4.8 ($>$500GeV) \\
Cas A & 6.9 & $<$0.66 ($>$500GeV)\\
%Rosette Nebula{\it $^3$} & 13.1 & $<$1.41 ($>$500GeV) \\
W44 & 6 & $<$3.0 ($>$300GeV) \\
W51 & 7.8 & $<$3.6 ($>$300GeV) \\
$\gamma$-Cygni & 9.3 & $<$2.2 ($>$300GeV) \\
W63 & 2.3 & $<$6.4 ($>$300GeV) \\
Tycho & 14.5 & $<$0.8 ($>$300GeV)\\
{\bf CAT} & & \\
CasA & 24.4 & $<$0.74 ($>$400GeV) \\\hline
\end{tabular*}

\begin{tablenotes}
{\it $^{a}$}{A different definition of Energy Threshold is used}

{\it $^{b}$}{Evidence for emission at the 4.9$\sigma$ level (P\"{u}hlhofer et al. 2001)}

{\it $^{c}$}{Limits converted from Crab units using flux of Hillas et al. 1998}
\end{tablenotes}
\end{table}

\begin{figure}[p]
\centerline{\resizebox{0.8\textwidth}{!}{\includegraphics*[]{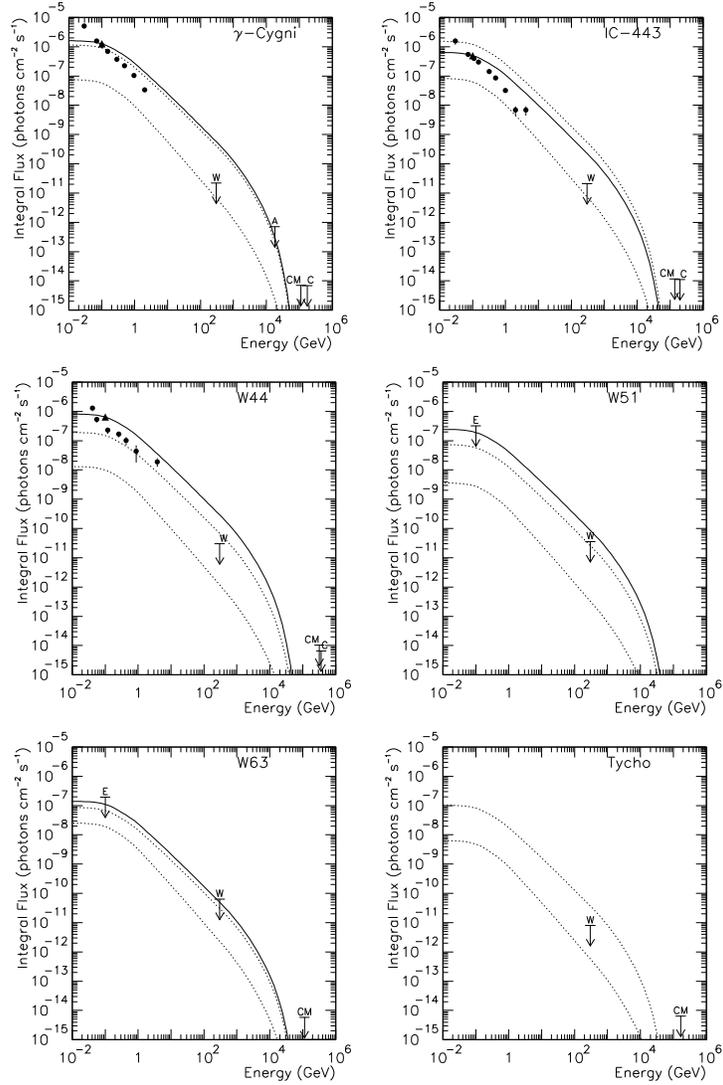}}}
\caption{\label{STSNR::BUCKLEY} MeV - TeV observations of shell-type 
SNR. Whipple upper limits marked as (W), associated EGRET fluxes as
points or EGRET upper-limits (E). Also shown are CASA-MIA (CM), Cygnus
(C) and AIROBICC upper-limits. The solid curve shows the extrapolation
of the DAV model from the EGRET integral data points at 100 MeV, shown
as a triangle. The dashed curve show the reasonable flux ranges of the
DAV model without making any assumptions about the EGRET data.}
\end{figure}

Recent observations of shell-type SNRs by the CANGAROO group have
resulted in the detections of SN1006 (Tanimori et al. 1998) and
RXJ1713.7-3946 (Muraishi et al. 2000). Observations of SN1006 in 1996
and 1997 show a significant excess from the NW rim of the SNR. The
excess is consistent with the location of non-thermal X-rays detected
by ASCA (Koyama et al. 1995). Similarly, RXJ1713.7-3946 has recently
had a ROSAT and ASCA X-ray source associated with it (Pfeffermann \&
Aschenbach 1996).  The TeV \Gray signal from these objects, if
confirmed, would not necessarily imply that they are accelerators of
hadronic cosmic-rays. It is considered likely that the \Grays are
linked through inverse-Compton scattering with a population of
relativistic electrons which are also responsible for producing the
X-rays. In any case, further observations and the determination of the
energy spectrum with more sensitive instruments will be required
either to confirm or to completely rule out the presence of a hadronic
component to the TeV signal.

The latest shell-type SNR to be detected at TeV energies is Cassiopeia
A, recently announced as a source by the HEGRA collaboration
(P\"{u}hlhofer et al. 2001). Based on 232 hours of observations they
report an excess at the 4.9$\sigma$ level, and calculate a flux of
$\mathrm{F}=5.8\pm1.2_{stat}\pm2_{syst}\times10^{-13}
\mathrm{cm}^{-2}\mathrm{s}^{-1}$ at (E$>$ 1 TeV).
Whether the \Grays from Cas A arise from inverse-Compton interactions
or through neutral pion decay has not yet been determined. There are
reasons to think that both components may be present at some
level. First, Cas A is associated with a bright source of hard
X-rays which indicates a population of non-thermal electrons with
energies up to 100 TeV (Allen et al. 1999). Second, Cas A is situated
in a region of high ambient matter density which has been associated
with the wind system left over from the progenitor. Measurement of the
energy spectrum will be required before any determination between the
two components can be made.

\section{Unidentified Sources}

IACTs are most sensitive when observing on-axis point sources,
i.e. when pointing directly towards an object of small angular
extent. Observations of extended sources or of sources where the
location is not well known are possible, however, as IACTs do have
good sensitivity across most of their field of view. It is therefore
possible to create a two dimensional VHE map of the sky within,
typically, $~0.5-1.0^\circ$ of the center of the field of view. The
angular resolution of a single IACT is $~0.15^\circ$ (Lessard et
al. 2000), and better for an array of IACTs, providing the ability to
resolve the location of a source within the field of view to high
accuracy.

VHE observations of unidentified EGRET sources are usually made on the
basis of their luminosity, spectrum and size of their error box or if
they have a good candidate association, such as an SNR or pulsar which
allows sensitive, point source observations to be made.

Results of previous observations of unidentified EGRET sources with
the Whipple 10m telescope (Buckley et al. 1997) and new results from
ongoing observations are given in table \ref{EUID::OBSTABLE}. No
unidentified sources have been detected in VHE \Grays but for many of
them, upper limits have been placed on VHE emission from within the
EGRET error circle.

\begin{table}[ht]
\caption[TeV Observations of Unidentified Sources]
{\label{EUID::OBSTABLE}TeV Observations of Unidentified Sources}
\begin{tabular*}{\textwidth}{@{\extracolsep{\fill}}lcc}\hline
{\bf Object} & {\bf Exposure time} & {\bf Flux Upper Limit} \\
{\bf Name} & {\bf (mins)} &{\bf  x $10^{-11}cm^{-2}s^{-1}$} \\\hline

J0241+6119 & 972 & 1.02{\it $^a$} \\
J0433+2907 & 499 & 4.4{\it $^a$} \\
J0545+3943 & 108 & 6.72{\it $^a$} \\
J0618+2234 & 1188 & 0.911{\it $^a$} \\
J0635+0521 & 108 & 5.59{\it $^a$} \\
J0749+17 & 486 & 0.813{\it $^a$} \\
J1746-2852 & 270 & 0.45{\it $^b$}\\
J1825-1307 & 702 & 1.55{\it $^a$} \\
J1857+0118 & 351 & 2.79{\it $^a$} \\
J2016+3657 & 287 & 5.8{\it $^a$} \\
J2020+4026 & 513 & 0.990{\it $^a$} \\
J2227+6122 & 274 & 6.0{\it $^a$} \\
\hline
\end{tabular*}

\begin{tablenotes}
{\it $^{a}$}Integral Flux Above 400 GeV.

{\it $^{b}$}Integral Flux Above 2.0 TeV.
\end{tablenotes}
\end{table}

\section{The Next Generation}

The future of ground-based \Gray astronomy lies with the next
generation of observatories that will build on the advances made to
date. Significant discoveries will come from extending the energy
range observable from the ground and by improving the flux
sensitivity.

Extending the observable energy range below 200GeV will be achieved by
building instruments with larger mirror area to gather more light from
the shower. Two approaches have been suggested. The first uses
existing solar furnace facilities which have fields of large
heliostats that lie unused at night. The arrival of the Cherenkov
wavefront at groups of heliostats is precisely measured and this
information is used to differentiate \Grays from cosmic-rays (Ong
1998). STACEE (Chantell et al. 1998), CELESTE (Quebert et al. 1995)
and Solar-2 (T\"{u}mer et al. 1999), each of which use this technique,
have started operating in the last few years and hope to observe at
$\sim$20-30 GeV. MAGIC (Barrio et al. 1998) takes a different
approach; by extending the size of a traditional IACT to 17m they plan
to achieve a significantly lower energy threshold than is possible
with current generation telescopes.

At the other end of the VHE spectrum, air shower arrays on very high
mountains, such as the Tibet Air Shower Array (Amenomori 1999) and
Cherenkov telescopes operating at large zenith angles have been joined
by MILAGRO (Sinnis et al. 1995), a water Cherenkov detector in New
Mexico that views all of the visible sky with $45^\circ$ of the zenith
and operates 24 hours a day and hopes to achieve a better energy
overlap with traditional ground-based techniques.

\begin{figure}[ht]
\centerline{\resizebox{!}{0.45\textheight}{\includegraphics[]{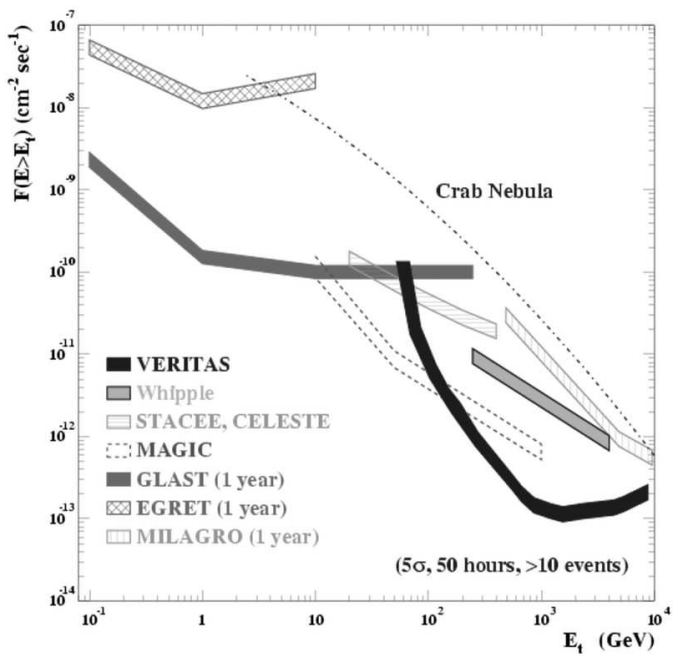}}}

\caption{\label{NXTGEN::SENSITIVITY}Point source sensitivity of some 
of the current and next generation of \Gray instruments; Whipple
(Weekes et al. 1989), MAGIC (Barrio et al. 1998), STACEE \& CELESTE
(Chantell et al. 1998 \& Quebert et al. 1995), HEGRA (Daum et
al. 1997), GLAST (Gehrels \& Michelson 1999), EGRET (Thompson et
al. 1993) and MILAGRO (Sinnis et al. 1995).}
\end{figure}

To significantly increase flux sensitivity and angular resolution
while achieving a broad operating energy range of 100GeV to 10TeV a
number of groups have undertaken to build arrays of large
($\sim10-12$m) IACTs operating in coincidence. VERITAS (Bradbury et
al. 1999), a system of seven 10m telescopes in Arizona will build on
the experience gained by the Whipple group. HESS (Hoffmann 1997), an
array of four (and possibly sixteen when completed) of 12m telescopes
in Namibia will extend the ideas pioneered by HEGRA. The
Japanese-Australian CANGAROO group are building an array of four 10m
telescopes at the site of the present CANGAROO experiment, one
telescope has been completed. Collectively, these three arrays and the
MAGIC telescope are often referred to as the Next Generation Gamma-Ray
Telescopes (NGGRTs).

Figure \ref{NXTGEN::SENSITIVITY} shows how all of these upcoming
experiments will overlap with the EGRET and Whipple sensitivities and
also with the sensitivity of the upcoming GLAST mission.  It is
evident that the NGGRTs will provide some overlap with GLAST and the
solar arrays, and at the high energy side with the Tibet Air Shower
array and MILAGRO.

The future of ground-based \Gray astronomy will be very
interesting. Continued cooperation with space-based missions will
ensure that the energy range from 1MeV to 10TeV will be well covered
in both the northern and southern hemispheres. Better flux sensitivity
will mean that new sources will be discovered; excellent angular
resolution will ensure that accurate source locations and associations
will be found. Detailed energy spectra in this range will allow for
better models of the emission regions.

\section{Acknowledgements}

Research in VHE \Gray astronomy at the Whipple Observatory is
supported by a grant from the U.S. Department of Energy. My thanks to
Deirdre Horan and Trevor Weekes for their helpful comments.

\end{document}